\documentclass[prb,letterpaper,twocolumn,aps,floatfix]{revtex4-2}
\usepackage{graphicx}
\usepackage{xcolor}
\usepackage{amsmath}
\usepackage{amsfonts}
\usepackage[small,bf]{subfigure}
\newcommand{\beq}{\begin{equation}}
\newcommand{\eeq}{\end{equation}}
\newcommand{\bk}{{{\bf{k}}}}

\newcommand{\br}{{{\bf{r}}}}

\newcommand{\bq}{{\bf{q}}}

\newcommand{\beqa}{\begin{eqnarray}}
\newcommand{\eeqa}{\end{eqnarray}}

\newcommand{\bsigma}{{\boldsymbol \sigma}}

\newcommand{\ra}{\rightarrow}

\newcommand{\cD}{{\cal D}}

\begin{document}
\title{Quantum geometric contribution to the diffusion constant}
\author{A.A. Burkov}
\affiliation{Department of Physics and Astronomy, University of Waterloo, Waterloo, Ontario 
N2L 3G1, Canada} 
\affiliation{Perimeter Institute for Theoretical Physics, Waterloo, Ontario N2L 2Y5, Canada}
\date{\today}
\begin{abstract}
We discuss the quantum geometric contribution to the diffusion constant and the DC conductivity in metals and semimetals with linear Dirac dispersion. 
We demonstrate that, for systems with perfectly linear dispersion, there exists a clear and rigorous separation of the quantum geometric from the ordinary band velocity 
contributions to the diffusion constant, which turns out to be directly related to the separation of a rank two tensor into transverse and longitudinal parts. 
We also demonstrate that, within the self-consistent Born approximation and for Gaussian-distributed disorder, the diffusion constant of three-dimensional Dirac fermions 
at charge neutrality is entirely quantum geometric in origin, which is not the case for two-dimensional Dirac fermions. This is the result of an accidental perfect cancellation 
of the band velocity contribution in three dimensions. 
\end{abstract}
\maketitle
\section{Introduction}
\label{sec:1}
There has been significant recent interest in quantum geometry, which in essence means nontrivial momentum dependence of the Bloch wavefunctions. 
The well-known Berry curvature is a paradigmatic quantum geometric quantity, which leads to the anomalous Hall effect and other phenomena, but the recent 
attention has instead been focused on the quantum metric, which has not been explored in as much detail in the past~\cite{Xie24,Bernevig25,Queiroz25,Nagaosa25}. 

In a typical metal with a large Fermi surface, most observable properties can be related to the band dispersion on the Fermi surface. 
For example, the DC conductivity is related to the Fermi surface average of the band velocity squared, while interaction-driven instabilities are 
determined by the nesting properties of the band dispersion at the Fermi energy. 
Quantum geometric effects also exist in this case, but appear as small corrections, which are typically ignored. 

However, there exist metallic systems without a well-defined Fermi surface, notably flat-band systems, such as two-dimensional (2D) systems 
in the quantum Hall regime or, more recently, moire materials~\cite{Bistritzer-MacDonald,Cao18-1,Cao18-2}. Another important example is topological semimetals, where the Fermi surface shrinks into lower-dimensional objects, i.e. discrete band-touching points or lines~\cite{Volovik03,Volovik07,Murakami07,Wan11,Burkov11-1,Burkov11-2}. 
In both cases the standard description is not applicable due to the lack of a regular Fermi surface and this is when quantum geometric effects may be expected to 
dominate~\cite{Torma15,Mitscherling20,Mitcherling22,Torma23,Law24,Savary25,Jin25,Roche25}.

In this paper we will be interested in one of the simplest properties of a metal, namely the DC electrical conductivity. 
For flat-band systems this has in fact been explored already in the classic context of transport in the lowest Landau level~\cite{Houghton82,Girvin82}, where the conductivity 
of a partially-filled Landau level is determined by the quantum geometry, although this terminology was not used at that time. 
Here we will instead focus on systems with band-touching points, specifically with Dirac dispersion in $d$-dimensions with $d \geq 2$. 
Instead of directly calculating the conductivity using Kubo formula~\cite{Mitcherling22,Torma23,Law24,Jin25}, we will consider the diffusion constant, related to the conductivity by the 
Einstein relation, which follows from charge conservation arguments. As will be seen below, this reveals some interesting 
properties, that are hidden in the Kubo formula for the DC conductivity, but are explicit when the diffusion constant is considered instead. 
In addition, the diffusion constant may be viewed as the stiffness of the diffusion modes, which are the Goldstone modes of the spontaneously broken 
retarded/advanced symmetry within the replica field theoretic formulation of the transport problem~\cite{Altland_Simons}. 
This connects the quantum geometric aspects of the DC transport problem to those of the superconductivity and magnetism in flat bands~\cite{Torma15}. 

We find that, in the case of Dirac fermions with perfectly linear dispersion, contributions to the diffusion constant may be cleanly separated into ``ordinary", 
which are related to the band velocity, and quantum geometric, which are not. 
This turns out to be directly related to the separation of a rank two tensor into longitudinal and transverse pieces, which exists in any rotationally-invariant system. 
In addition, we demonstrate that the diffusion constant of 3D Dirac fermions at charge neutrality has a purely quantum geometric origin, 
with all ordinary band velocity contributions cancelling out [within the self-consistent Born approximation (SCBA) and for Gaussian-distributed disorder]. This cancellation only happens in 3D and is accidental, i.e. not related to any physical properties of the 3D Dirac fermions. 
We will also relate our results to the previous calculations of the DC conductivities for 2D and 3D Dirac fermions~\cite{Fradkin86-1,Fradkin86-2,Ludwig94,Ziegler97,Ando98,Shovkovy02,CastroNeto06,Mirlin06,Ludwig07,Altland16}. 
For recent experimental context see e.g. Ref.~\cite{Ye25}. 

The rest of the paper is organized as follows. 
In Section~\ref{sec:2} we derive the diffusion propagator for d-dimensional Dirac fermions, both in the heavily-doped case with a large Fermi energy and at charge neutrality. 
We start by reviewing the doped case, which corresponds to an ordinary metal with a large Fermi surface. We then discuss the case of a $d$-dimensional semimetal 
at charge neutrality, in which case the Fermi surface is absent (shrinks to a point) and derive an expression for the diffusion constant, in which band velocity and quantum geometric 
contributions are clearly separated, which is our main new result. 
We conclude in Section~\ref{sec:3} with a brief discussion of our results and possible directions for future work. 
\section{Diffusion propagator for massless Dirac fermions}
\label{sec:2}
We will be concerned with $d$-dimensional massless Dirac fermions, where $d \geq 2$, since the $d=1$ massless Dirac fermion simply corresponds to a regular 1D single-band metal, 
which does not have nontrivial quantum geometry. 
It will prove useful to keep the space dimensionality $d$ general in some of the calculations, but we will specialize to physical $d = 2, 3$ dimensions where necessary.
 
We will base our discussion on massless nondegenerate Dirac (Weyl) fermions in physical dimensions $d = 2, 3$, described by the Hamiltonian
\beq
\label{eq:1}
H = v_F \bsigma \cdot \bk, 
\eeq
where $\sigma$ are Pauli matrices, $\bk$ is the crystal momentum with appropriate number of components and $v_F$ is a constant Fermi velocity. 
$\hbar = c = 1$ units will be used throughout, except in some of the final results. 
The energy eigenvalues are given by 
\beq
\label{eq:2}
\epsilon_{s \bk} = s v_F k \equiv s \epsilon_\bk, 
\eeq
where $s = \pm $. 
The corresponding eigenvectors are 
\beq
\label{eq:3}
|u^s_\bk \rangle = \frac{1}{\sqrt{2}} \left(1, s e^{i \varphi} \right)^T, 
\eeq
where $e^{i \varphi} = (k_x + i k_y)/k$, for $d = 2$.
For $d = 3$ the eigenvectors are given by
\beq
\label{eq:4}
|u^s_\bk \rangle = \frac{1}{\sqrt{2}} \left(\sqrt{1 + s \frac{k_z}{k}}, s e^{i \varphi} \sqrt{1 - s \frac{k_z}{k}} \right)^T, 
\eeq
where $e^{i \varphi} = (k_x + i k_y)/\sqrt{k_x^2 + k_y^2}$. 

We will assume Gaussian-distributed scalar disorder potential, characterized by the correlator 
\beq
\label{eq:5}
\langle V(\br) V(\br') \rangle = \gamma^2 \delta(\br - \br'). 
\eeq
The variance $\gamma$ characterizes the width of the gaussian distribution of $V(\br)$, i.e. the strength of the disorder. 
\subsection{d-dimensional metal}
\label{sec:2.1}
\begin{figure}[t]
\includegraphics[width=9cm]{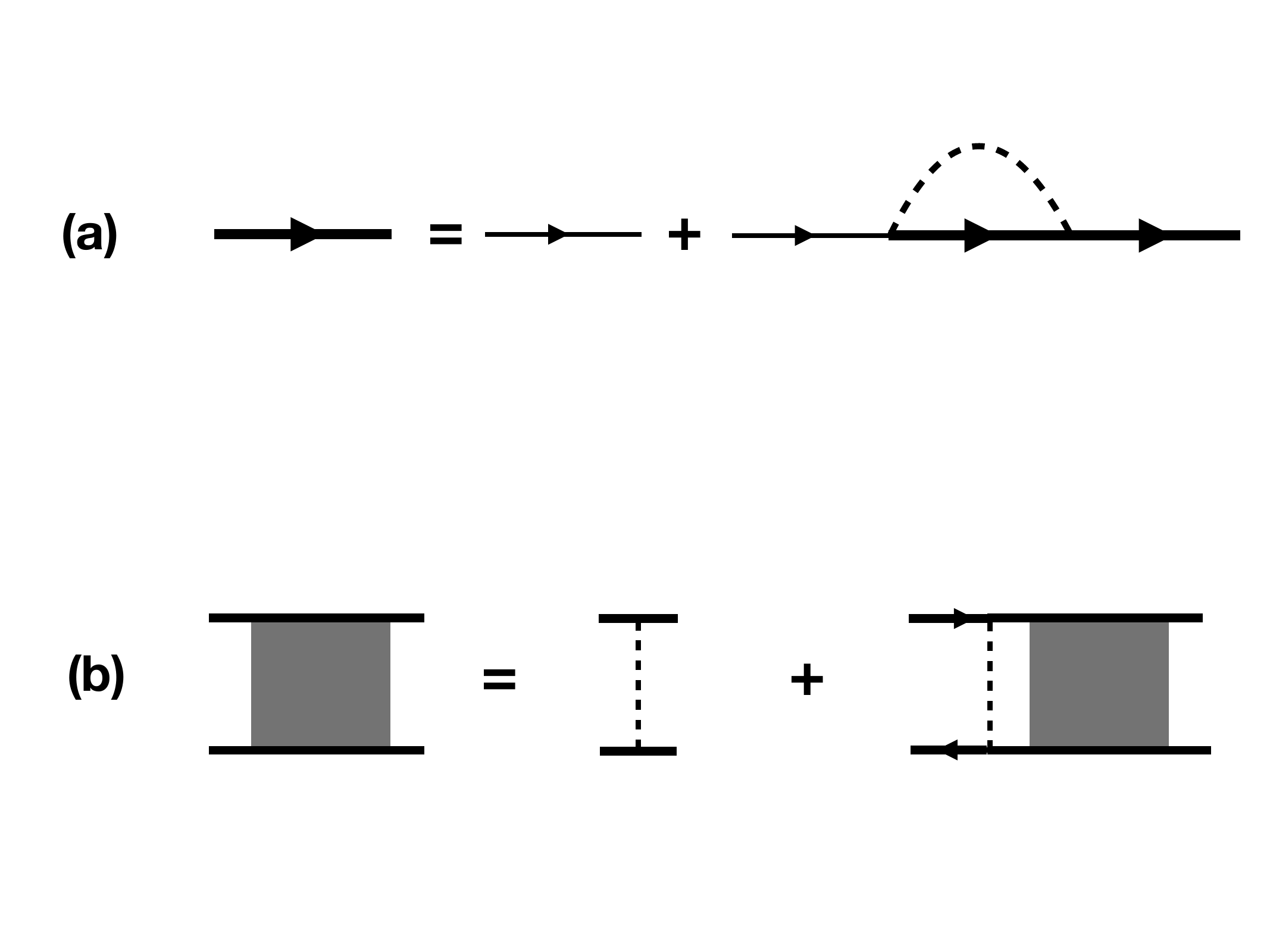}
\vspace{-1cm}
\caption{Diagrammatic representation of (a) The SCBA Green's function. Thin line represents the bare Green's function, thick line is the SCBA impurity-averaged Green's function and the dashed line represents the disorder potential correlator $\langle V(\br) V(\br') \rangle = \gamma^2 \delta(\br - \br')$. (b) The equation for the diffusion propagator $\cD$.}
\label{fig:1}
\end{figure}

Let us start by reviewing the derivation of the propagator of diffusion modes, from which the diffusion constant and the DC conductivity may be read off, for the standard case 
of a Fermi liquid with a large Fermi surface. 
In the present context this corresponds to a heavily-doped system, such that the Fermi energy $\epsilon_F$ satisfies $\epsilon_F \gg 1/\tau$, where $\tau$ is the momentum relaxation time, 
to be calculated below. 
When $\epsilon_F \tau \gg 1$, transport properties may be calculated using the SCBA, along with summing ladder impurity-averaging diagrams 
for the diffusion propagator, see Fig.~\ref{fig:1}. 
The approximation here consists of ignoring all diagrams with intersecting impurity lines, which may be shown to be small, compared to the non-crossing diagrams, the small 
parameter being $1/\epsilon_F \tau$, or equivalently $1/ k_F \ell$, where $k_F$ is the Fermi momentum and $\ell = v_F \tau$ is the mean free path~\cite{Altland_Simons}. 
This approximation is conceptually equivalent to the saddle-point (Hartree-Fock) plus quadratic fluctuations (RPA) approximation in the interacting electron problem, which may be seen explicitly in the replica field theory formulation~\cite{Altland_Simons}. We will stick to the diagrammatic language here, since it is more straightforward for what we need. 

The SCBA equation for the retarded disorder-scattering self-energy is given by
\beq
\label{eq:6}
\Sigma^R_s(\bk, \omega) = \gamma^2 \int \frac{d^d k'}{(2 \pi)^d} \sum_{s'} | \langle u^s_\bk | u^{s'}_{\bk'} \rangle |^2 G^R_{s'}(\bk', \omega), 
\eeq
where the retarded disorder-averaged Green's function in the Dirac band basis is  
\beq
\label{eq:7}
G^R_s(\bk, \omega) = \frac{1}{\omega + \epsilon_F - s \epsilon_{\bk} - \Sigma^R_s(\bk, \omega)}. 
\eeq

Let us assume for concreteness that $\epsilon_F > 0$. In this case, when $\epsilon_F \tau \gg 1$, we may ignore the contribution of the valence band and 
look for the solution of Eq.~\eqref{eq:6} in the form
\beq
\label{eq:8} 
\textrm{Im} \Sigma^R_+(\bk, \omega) = - \frac{1}{2 \tau}, 
\eeq
assuming the frequency $\omega$ is small. 
The real part of the self-energy may be absorbed into $\epsilon_F$ and is of no interest. 
Then we obtain a self-consistent equation for the momentum relaxation rate
\beq
\label{eq:9}
\frac{1}{2 \tau} = \frac{\gamma^2}{2} \int \frac{d^d k}{(2 \pi)^d} \frac{\frac{1}{2 \tau}}{(\epsilon_\bk - \epsilon_F)^2 + \frac{1}{4 \tau^2}}, 
\eeq
where the factor of $1/2$ in front of the integral arises from the angular integral of the matrix element $| \langle u^s_\bk | u^{s'}_{\bk'} \rangle |^2$. 

In the limit $\epsilon_F \tau \gg 1$ this equation may be easily solved noticing that the density of states at the Fermi energy may be written as
\beqa
\label{eq:10}
g_d&=&\frac{S_d}{(2 \pi)^d} \epsilon_F^{d-1} = - \frac{1}{\pi} \int \frac{d^d k}{(2 \pi)^d} \textrm{Im} \frac{1}{\epsilon_F - \epsilon_\bk + i \eta} \nonumber \\
&\approx&\frac{1}{\pi} \int \frac{d^d k}{(2 \pi)^d} \frac{\frac{1}{2 \tau}}{(\epsilon_\bk - \epsilon_F)^2 + \frac{1}{4 \tau^2}}, 
\eeqa
where $\eta = 0+$ and 
\beq
\label{eq:11}
S_d = \int d \Omega_d = \frac{2 \pi^{d/2}}{\Gamma(d/2)},
\eeq
is the surface area of a unit sphere in $d$-dimensions.  
Substituting this into Eq.~\eqref{eq:9} gives the SCBA scattering rate
\beq
\label{eq:12}
\frac{1}{\tau} = \pi \gamma^2 g_d. 
\eeq
We will omit the $d$ index from the scattering time $\tau$ to avoid clutter. 

We may now proceed to calculate the diffusion propagator. 
In the same approximation as the SCBA for disorder self-energy, the diffusion propagator is given by the sum of ladder diagrams, shown in Fig.~\ref{fig:1}, 
which corresponds to a geometric sum
\beq
\label{eq:13}
\cD(\bq, \Omega) = \frac{1}{1 - I(\bq, \Omega)}, 
\eeq
where 
\beq
\label{eq:14}
I(\bq, \Omega) = \frac{\gamma^2}{2} \int \frac{d^d k}{(2 \pi)^d}\textrm{Tr} [G^R \left(\bk + \frac{\bq}{2}, \Omega\right) G^A\left(\bk - \frac{\bq}{2}, 0\right)], 
\eeq
and 
\beqa
\label{eq:15}
G^{R,A}(\bk, \omega)&=&\sum_s \frac{|u^s_\bk \rangle \langle u^s_\bk|}{\omega + \epsilon_F - s \epsilon_\bk \pm \frac{i}{2 \tau_s}} \nonumber \\
&=&\sum_s  |u^s_\bk \rangle \langle u^s_\bk| G^{R, A}_s(\bk, \omega). 
\eeqa
Here $1/\tau_+ = 1/\tau$ and $1/\tau_- = 0+$. 
This gives
\beqa
\label{eq:16}
I(\bq, \Omega)&=&\frac{\gamma^2}{2} \sum_{s, s'} \int \frac{d^d k}{(2 \pi)^d} |\langle u^s_{\bk + \bq/2} | u^{s'}_{\bk - \bq/2} \rangle|^2 \nonumber \\
&\times&G^R_s\left(\bk + \frac{\bq}{2}, \Omega\right) G^A_{s'}\left(\bk - \frac{\bq}{2}, 0\right). 
\eeqa
We will be interested in small $q$ and $\Omega$ behavior of the function $I(\bq, \Omega)$. 
It follows from general, independent of microscopic details, considerations of charge conservation, that the Taylor expansion of this function has the form~\cite{Altland_Simons}
\beq
\label{eq:17}
I(\bq, \Omega) \approx 1 + i \Omega \tau - D q^2 \tau, 
\eeq
where $D$ is, by definition,  the diffusion constant. 
We are interested specifically in calculating $D$, which, according to Eq.~\eqref{eq:17}, can be found by expanding the function $I(\bq, 0)$ to second order in $q$. 

We can see from Eq.~\eqref{eq:16} that the $\bq$-dependence arises from two distinct sources: the wavefunction overlap $\langle u^s_{\bk + \bq/2} | u^{s'}_{\bk - \bq/2} \rangle$ 
and the $\bq$-dependence of the retarded and advanced Green's functions in the band eigenstate basis, which comes from the momentum dependence of the band dispersions. 
The wavefunction overlap may be expressed in terms of the quantum geometric tensor as
\beq
\label{eq:18}
|\langle u^s_{\bk + \bq/2} | u^{s'}_{\bk - \bq/2} \rangle|^2 \approx \delta_{s s'} - g_{a b}^{s s'}(\bk) q_a q_b, 
\eeq
where repeated $a, b$ indices are summed over and the geometric tensor $g^{s s'}_{a b}(\bk)$ is given by
\beq
\label{eq:19}
g^{s s'}_{a b}(\bk) = \delta_{s s'} \textrm{Re}(\langle \partial_a u^s_\bk | \partial_b u^s_\bk \rangle) - A^{s s'}_a(\bk) A^{s' s}_b(\bk), 
\eeq
where $A^{s s'}_a(\bk) = - i \langle u^s_\bk | \partial_a u^{s'}_\bk \rangle$ is the nonabelian Berry connection. 
A straightforward calculation gives for massless Dirac fermions in any dimension $d \geq 2$
\beq
\label{eq:20}
g^{s s'}_{a b}(\bk) = \frac{2 \delta_{s s'}-1}{4 k^2}(\delta_{a b} - \hat k_a \hat k_b) \equiv (2 \delta_{s s'} - 1) g_{a b}(\bk).
\eeq
Note that $g_{a b}(\bk)$ is a purely transverse rank two tensor, which expresses the fact that it 
measures distances on the Bloch sphere, defined by the unit vector $\hat k$. This will play an important role in what follows.

The second part of the $\bq$-dependence in Eq.~\eqref{eq:16} arises from the product of the retarded and advanced Green's functions. 
To better reveal its physical meaning, it is useful to rewrite this product as
\beq
\label{eq:21}
G^R G^A = \frac{G^A - G^R}{(G^R)^{-1} - (G^A)^{-1}}. 
\eeq
The point of this is that it re-expresses the $G^R G^A$ product in terms of more physically-meaningful pieces: $G^A - G^R$ factor is related to the spectral weight, 
while $(G^R)^{-1} - (G^A)^{-1}$ is essentially the inverse of the particle-hole pair propagator, as will be explained in greater detail below. 
This gives
\beqa
\label{eq:22}
&&G^R_s\left(\bk + \frac{\bq}{2}, 0 \right) G^A_{s'}\left(\bk - \frac{\bq}{2}, 0\right) = \nonumber \\
&=&\frac{G^R_{s}\left(\bk + \frac{\bq}{2}, 0\right) - G^A_{s'}\left(\bk - \frac{\bq}{2}, 0 \right)}
{s \epsilon_{\bk + \frac{\bq}{2}} - s' \epsilon_{\bk - \frac{\bq}{2}} - \frac{i}{ 2 \tau_s} - \frac{i}{2 \tau_{s'}}}. 
\eeqa

Keeping in mind that we are only interested in the $\bq$-dependence up to second order, we may rewrite the function $I(\bq, 0)$ as a sum of two terms 
\beq
\label{eq:23}
I(\bq, 0) = I_1(\bq) + I_2(\bq), 
\eeq
where 
\beq
\label{eq:24} 
I_1(\bq) = \frac{\gamma^2}{2} \sum_s \int \frac{d^d k}{(2 \pi)^d} \frac{G^R_s \left(\bk + \frac{\bq}{2}, 0\right) - G^A_s\left(\bk - \frac{\bq}{2}, 0 \right)}{s \epsilon_{\bk + \frac{\bq}{2}} - s \epsilon_{\bk - \frac{\bq}{2}} - \frac{i}{\tau_s}},
\eeq
and 
\beqa
\label{eq:25}
&&I_2(\bq) = - \frac{\gamma^2}{2} \sum_{s, s'} \int \frac{d^d k}{(2 \pi)^d} \nonumber \\
&\times&\frac{g^{s s'}_{a b}(\bk) q_a q_b}{\left(\epsilon_F - s \epsilon_\bk + \frac{i}{2 \tau_s}\right)\left(\epsilon_F - s' \epsilon_\bk - \frac{i}{2 \tau_{s'}} \right)}.
\eeqa

The second term $I_2(\bq)$ is entirely quantum geometric in origin and contains both inter- and intra-band contributions. 
The first term $I_1(\bq)$ is intra-band only and, as we will see below, contains both band velocity and geometric contributions. 
Let us define
\beq
\label{eq:26}
\Delta G_s(\bk, \bq) = G^R_s \left(\bk + \frac{\bq}{2}, 0\right) - G^A_s\left(\bk - \frac{\bq}{2}, 0 \right). 
\eeq
The $\bq = 0$ limit of this quantity has a simple physical meaning, namely 
\beq
\label{eq:27}
i \Delta G_s(\bk, 0) = \varrho_s(\bk), 
\eeq
where $\varrho_s(\bk)$ is the disorder-broadened spectral density, which gives the density of states at the Fermi energy in band $s$, when integrated over $\bk$. 
At finite $\bq$, $\Delta G_s(\bk, \bq)$ does not have such a simple physical meaning any longer, however, loosely, it is the spectral density of particle-hole pairs, with particle carrying momentum $\bk + \bq/2$ and hole $\bk - \bq/2$. 
The denominator in Eq.~\eqref{eq:22} is the inverse propagator of a particle-hole pair in band $s$, with the particle having momentum $\bk + \bq/2$ and the hole $\bk - \bq/2$. 
Such propagating particle-hole pairs, integrated over the center-of-mass momentum $\bk$ and weighted by the function $\Delta G_s(\bk, \bq)$, correspond 
to density fluctuations at wavevector $\bq$, which relax via diffusion. 

At this point, let us first specialize to the case of an ordinary ``good" metal with $\epsilon_F \tau \gg 1$. 
In this case, it is easy to see that the intra-band contribution, corresponding to the conduction band $s = +$ in $I_1(\bq)$, dominates. 
All other contributions, including the entire quantum-geometric term $I_2(\bq)$, are higher-order in the small parameter $1/\epsilon_F \tau$ and may be discarded. 
In fact, they should be discarded since SCBA itself consists of neglecting quantum interference contributions, which 
correspond to diagrams with crossed impurity lines and which are also of higher order in $1/\epsilon_F \tau$. 
The $\bq$-dependence of $\Delta G_s(\bk, \bq)$ may also be ignored in this case, since it will only produce a correction to the uniform density of states at the Fermi energy, 
small in the same parameter $1/\epsilon_F \tau$. 
Then we obtain
\beq
\label{eq:28}
I(\bq, 0) \approx I_1(\bq) \approx - \frac{i \gamma^2}{2} \int \frac{d^d k}{(2 \pi)^d} \frac{\varrho_+(\bk)}{v_F \bq \cdot \hat k - \frac{i}{\tau}}, 
\eeq
where we have expanded the denominator to second order in $\bq$ (the quadratic term vanishes).
Expanding to second order in $\bq$ we obtain
\beq
\label{eq:29}
I(\bq, 0) \approx \frac{\gamma^2 \tau}{2} \int \frac{d^d k}{(2 \pi)^d} \varrho_+(\bk) [1 - v_F^2 \tau^2 (\bq \cdot \hat k)^2], 
\eeq
where the linear term vanishes upon integration over $\bk$ and has been discarded.  
Using 
\beq
\label{eq:30}
\int \frac{d^d k}{(2 \pi)^d} \varrho_+(\bk) = 2 \pi g_d, 
\eeq
and 
\beq
\label{eq:31}
\langle \hat k_a \hat k_b \rangle_d \equiv \frac{1}{S_d} \int d \Omega_d\, \hat k_a \hat k_b = \frac{1}{d} \delta_{a b}, 
\eeq
where $\langle \ldots \rangle_d$ denote directional average over the surface of a $d$-dimensional unit sphere, we obtain
\beq
\label{eq:32}
I(\bq, 0) \approx 1 - \frac{v_F^2 \tau^2}{d} \bq^2. 
\eeq
Thus we have obtained the standard expression for the diffusion constant of a $d$-dimensional metal
\beq
\label{eq:33}
D_d = \frac{v_F^2 \tau}{d}. 
\eeq
The DC conductivity may be obtained from the Einstein relation
\beq
\label{eq:34}
\sigma_d = e^2 g_d D_d = \frac{S_d}{d (2 \pi)^{d-1}} \frac{e^2}{h} k_F^{d -2} k_F \ell, 
\eeq
where we have restored explicit Planck's constant. 
This is the standard result for the conductivity of a normal metal: the conductivity is large, namely it is given by the conductance quantum $e^2/h$, multiplied 
by a large number $k_F \ell$ and the appropriate dimensional factor $k_F^{d - 2}$. 

\subsection{Dirac semimetal}
\label{sec:2.2}
Now let us consider the case of a Dirac semimetal, with $\epsilon_F \ra 0$. 
In this case both bands contribute to the SCBA equation Eq.~\eqref{eq:6} equally and the solution is 
\beq
\label{eq:35}
\textrm{Im} \Sigma^R_+(\bk, \omega) =  \textrm{Im} \Sigma^R_-(\bk, \omega) =  - \frac{1}{2 \tau}, 
\eeq
where the scattering rate satisfies the equation
\beq
\label{eq:36}
\frac{1}{2 \tau} = \frac{\gamma^2}{2 \tau} \int \frac{d^d k}{(2 \pi)^d} \frac{1}{\epsilon_\bk^2 + \frac{1}{4 \tau^2}}. 
\eeq
The integral on the right-hand-side requires an upper momentum cutoff $\Lambda$, which is of the order of the primitive reciprocal lattice vector. 
The solution is strongly dimension-dependent and we will therefore consider the $d = 2, 3$ cases separately. 
In the $d = 2$ case, we obtain
\beq
\label{eq:37}
1 = \frac{\gamma^2}{4 \pi v_F^2} \ln(1 + 4 \Lambda^2 \ell^2). 
\eeq
In the limit $\Lambda \ell \gg 1$, the solution is
\beq
\label{eq:38}
\frac{1}{\tau} = v_F \Lambda e^{- 2 \pi v_F^2/\gamma^2},  
\eeq
which corresponds to the well-known result that in a 2D Dirac semimetal even infinitesimally small amount of disorder generates a finite, albeit exponentially small, 
density of states~\cite{Ando98}. 

For $d = 3$ we get
\beq
\label{eq:39}
1 = \frac{\gamma^2 \Lambda}{2 \pi^2 v_F^2} \left[1 - \frac{1}{2 \Lambda \ell} \textrm{arctan}(2 \Lambda \ell) \right]. 
\eeq
Unlike in 2D, this equation has a nontrivial solution only for $\gamma > \gamma_c$, where 
\beq
\label{eq:40}
\gamma_c = \sqrt{\frac{2 \pi^2 v_F^2}{\Lambda}}. 
\eeq
Near $\gamma_c$, the solution behaves as
\beq
\label{eq:41}
\frac{1}{\tau} \approx \frac{8 v_F \Lambda}{\pi} \frac{\gamma - \gamma_c}{\gamma_c}, 
\eeq
i.e. the scattering rate vanishes linearly with $\gamma - \gamma_c$. 
The SCBA thus predicts a phase transition at a finite strength of disorder from a clean 3D Weyl or Dirac semimetal, in which disorder is irrelevant when $\gamma < \gamma_c$, to
a diffusive metal with a finite disorder-induced density of states when $\gamma > \gamma_c$. 
The validity of this result is known to be questionable due to nonperturbative rare region effects, which likely lead to a finite density of states and diffusive transport at arbitrarily weak scalar disorder~\cite{Nandkishore14,Pixley15,Pixley16,Yi25}.
However, this subtlety is irrelevant for the issues that we will be discussing here and we will proceed assuming $\gamma > \gamma_c$, while keeping in mind that in reality 
$\gamma_c$ may be zero. 

The diffusion propagator still has the same form of Eqs.~\eqref{eq:14} and \eqref{eq:23}-\eqref{eq:25}, but there is no longer a separation into standard band-velocity terms, which are dominant, and corrections, which are small in the parameter $1/\epsilon_F \tau$ and may be discarded. 
All contributions are now of the same order and must be included. In principle, even the validity of the SCBA itself is now questionable, since it also relies on $\epsilon_F \tau \gg 1$. 
Nevertheless, quantum interference corrections may be separated from the rest, for example due to their distinct temperature dependence~\cite{Lee85}. 
As long as the temperature is not so low that the sample size is smaller than the phase coherence length (mesoscopic regime), quantum interference effects may be ignored. 
We will thus proceed to use SCBA and ladder approximation, with the expectation that even at low temperatures they produce at least qualitatively correct results. 

To obtain the correct result for the diffusion constant we now need to carefully expand all contributions to the inverse diffusion propagator $1 - I(\bq, 0)$ to second order in $\bq$. 
Compared to the case of a good metal with $\epsilon_F \tau \gg 1$, new contributions arise from the quantum geometric tensor, as well as from the $\bq$-dependence of the 
function $\Delta G_s(\bk, \bq)$. 
It will turn out to be useful to separate all the contributions into two groups, according to their dependence on the direction of the wavevector $\bk$, namely into 
longitudinal ones, proportional to $\hat k_a \hat k_b$ and transverse, proportional to $\delta_{a b} - \hat k_a \hat k_b$. 
As we have already seen above [see e.g. Eq.~\eqref{eq:29}], longitudinal contributions arise from the band velocity terms. 
In contrast, transverse contributions are associated with the quantum geometry, since the quantum geometric tensor is purely transverse, see Eq.~\eqref{eq:20}. 

Aside from the quantum geometric tensor in the function $I_2(\bq)$, the rest of the $\bq$-dependence is contained in the band energies in $I_1(\bq)$. 
Expanding them to second order, we have
\beq
\label{eq:42}
\epsilon_{\bk \pm \frac{\bq}{2}} \approx \epsilon_\bk \pm \frac{1}{2} v_F (\bq \cdot \hat k) + \frac{1}{8} q_a q_b \frac{\partial^2 \epsilon_\bk}{\partial k_a \partial k_b}.
\eeq
The last term involves the inverse effective mass tensor, which does not contribute to the inverse particle-hole pair propagator in the denominator of Eq.~\eqref{eq:24}, but does 
contribute to the $\bq$-dependence of the function $\Delta G_s(\bk, \bq)$ in the numerator. 
Importantly, the following relation holds for Dirac fermions in any dimension
\beq
\label{eq:43}
\frac{\partial^2 \epsilon_\bk}{\partial k_a \partial k_b} = 4 \epsilon_\bk g_{a b}(\bk), 
\eeq
i.e. the inverse effective mass tensor is proportional to the quantum geometric tensor. 
This relation is easy to understand. Since 
\beq
\label{eq:44} 
\frac{\partial \epsilon_\bk}{\partial k_a} = v_F \hat k_a, 
\eeq
the second derivative, being the derivative of a unit vector, will be purely transverse to $\hat k$ and thus must be proportional to the metric tensor. 
Thus the inverse effective mass tensor is the second source of quantum geometric contribution to transport properties, in addition to the explicit appearance of the quantum 
geometric tensor in Eq.~\eqref{eq:25}. 
One may worry that a direct relation between the inverse effective mass tensor and the quantum geometric tensor in Eq.~\eqref{eq:43} is a special property of Dirac fermions. 
Outside of the Dirac fermion context, the inverse effective mass tensor may appear to be an ``ordinary" band dispersion property, like band velocity, and thus its general 
interpretation as a quantum geometric quantity may seem to be questionable.  
Indeed, it may be defined even for a single parabolic band, which does not have any nontrivial quantum geometry. Note, however, that in this case the effective mass tensor also does not 
contribute to the diffusion constant, only the band velocity does, see Eq.~\eqref{eq:28}. 
In fact, a close relation between the effective mass tensor and quantum geometry is a general property of all multiband systems~\cite{Queiroz25}, 
it is just particularly straightforward for Dirac fermions. Dirac fermions may be viewed as a paradigm of metallic systems without a Fermi surface and we do expect that, at a qualitative level, our results will serve as a useful demonstration of the general role quantum geometry can play in dissipative transport in such systems.

Expanding $I(\bq, 0)$ to second order in $\bq$, we obtain, after straightforward algebra
\beqa
\label{eq:45}
&&I(\bq, 0) \approx 1 - \frac{\gamma^2 v_F^2}{4} q_a q_b \int \frac{d^d k}{(2 \pi)^d} \hat k_a \hat k_b \frac{\frac{3}{4 \tau^2} - \epsilon_\bk^2}{\left(\epsilon_\bk^2 + \frac{1}{4 \tau^2}\right)^3} 
\nonumber \\
&-&3 \gamma^2 q_a q_b \int \frac{d^d k}{(2 \pi)^d} g_{a b}(\bk) \frac{\epsilon_\bk^2}{\left(\epsilon_\bk^2 + \frac{1}{4 \tau^2} \right)^2}. 
\eeqa
Importantly, there are two distinct contributions: longitudinal, where the integrand is proportional to $\hat k_a \hat k_b$, and transverse, where 
the integrand is proportional to the quantum metric tensor $g_{a b}(\bk) \propto \delta_{a b} - \hat k_a \hat k_b$. 
The longitudinal contribution arises from the band velocity, hence the $v_F^2$ prefactor. It is not, however, the same as the band velocity contribution to the diffusion 
constant of an ordinary metal Eq.~\eqref{eq:29}, it is strongly renormalized by the $\bq$-dependent corrections to the spectral weight Eq.~\eqref{eq:26}. 
In fact, the renormalization is so strong that the Fermi velocity dependence completely drops out of the final result, as we will see below. 

Before explicitly evaluating the momentum integrals in Eq.~\eqref{eq:45}, it is useful to simplify the expression further. 
Using the identity
\beq
\label{eq:46}
k^{d-1} \frac{\frac{3}{4 \ell^2} - k^2}{\left(k^2 + \frac{1}{4 \ell^2}\right)^3} = \frac{d}{d k} \left[\frac{k^d}{\left(k^2 + \frac{1}{4 \ell^2}\right)^2}\right] + \frac{(3 - d) k^{d-1}}{\left(k^2 + \frac{1}{4 \ell^2}\right)^2}, 
\eeq
along with 
\beq
\label{eq:47}
\left.\frac{k^d}{\left(k^2 + \frac{1}{4 \ell^2}\right)^2}\right|_0^{\infty} = 0, 
\eeq
when $d < 4$ and assuming $\Lambda \ell \gg 1$, we may rewrite Eq.~\eqref{eq:45} as
\beq
\label{eq:48}
I(\bq, 0) \approx 1 - D_{a b} q_a q_b \tau, 
\eeq
which is the analog of Eq.~\eqref{eq:17}. 
The ``diffusion tensor" $D_{a b}$ is given by 
\beq
\label{eq:49}
D_{a b} = \frac{A_d \gamma^2 \tau^{3 - d}}{v_F^{d -2}} \left[\frac{3 - d}{3} \langle \hat k_a \hat k_b \rangle_d + \langle \delta_{a b} - \hat k_a \hat k_b \rangle_d \right], 
\eeq
where $A_d$ is a dimension-dependent numerical coefficient
\beq
\label{eq:50}
A_d = \frac{3 \pi S_d (2 - d)}{(4 \pi)^d \sin(\pi d/2)}. 
\eeq

Eq.~\eqref{eq:49} is our main result. We have intentionally kept the angular averages $\langle \hat k_a \hat k_b \rangle_d$ unevaluated, since this allows one to explicitly see distinct 
contributions to the diffusion constant: band velocity, which arises from the longitudinal tensor $\hat k_a \hat k_b$, and geometric, coming from the transverse tensor 
$\delta_{a b} - \hat k_a \hat k_b$. 
Note that the coefficient $A_d$ diverges as $d \ra 4$. This is because for $d \geq 4$, the momentum integrals in the expression for the 
inverse diffusion propagator Eq.~\eqref{eq:45} become cutoff-dependent and diverge as we take the limit $\Lambda \ell \ra \infty$. 
The total derivative term Eq.~\eqref{eq:47} also no longer vanishes in this case and our final result Eq.~\eqref{eq:49} is therefore invalid
and needs to be reconsidered. We will not further dwell on the $d \geq 4$ case here since we are interested in what happens in physical $d = 2, 3$ dimensions. 

One of the interesting features of Eq.~\eqref{eq:49} is that the longitudinal contribution to the diffusion constant apparently vanishes identically in 3D due to the $3 - d$ prefactor, 
which can be traced back to the identity Eq.~\eqref{eq:46}. 
This cancellation appears to be accidental, i.e. not related to any special physical properties of 3D Dirac fermions. 
Physically, this happens because two competing tendencies are present in the $\bq$-dependence of the longitudinal contribution: a positive contribution coming from the 
denominator in Eq.~\eqref{eq:24} (inverse particle-hole propagator) and negative one coming from the numerator, i.e. the spectral function $\Delta G_s(\bk, \bq)$. 
In high enough dimension the negative contribution wins and a sign change of the longitudinal contribution happens at $d = 3$. Note that the total diffusion constant always remains 
positive, as it should. 
As a result of this, the diffusion constant of 3D Dirac/Weyl fermions at charge neutrality has a purely quantum geometric origin, which is surprising. 
Note, that one certainly should not overestimate the significance of this result. Since the cancellation of the band velocity contribution is accidental, we do not expect it to be robust to various details, such as the specific model of the disorder potential, or corrections to SCBA, which is not particularly well justified in the absence of a Fermi surface. 
As already mentioned above, our results should mostly be viewed as a conceptual demonstration of the significant role quantum geometry may be expected to play in ordinary dissipative linear transport in metallic systems without a Fermi surface. 

Evaluating the angular averages using Eq.~\eqref{eq:31}, we obtain 
\beq
\label{eq:51}
D_{a b} = D_d \delta_{a b}, 
\eeq
where 
\beq
\label{eq:52}
D_d = \frac{2 A_d \gamma^2 \tau^{3 -d}}{v_F^{d -2}}, 
\eeq
is the diffusion constant of $d$-dimensional Dirac fermions. 
Another interesting feature of Eq.~\eqref{eq:49} is that, according to it, the ratio of the longitudinal (band velocity) to the transverse (geometric) contributions to the diffusion constant 
$D_d = D^{||}_d + D^{\perp}_d$ of Dirac fermions at charge neutrality is a universal, and very simple, function of dimensionality, given by
\beq
\label{eq:53}
\frac{D^{||}_d}{D^{\perp}_d} = \frac{3 - d}{3(d - 1)}.
\eeq
Assuming no cutoff dependence, the only dimensionless ratio one can construct out of the available physical quantities is $\gamma^2 \tau^{2 - d}/v_F^d$. 
In principle, the ratio $D^{||}_d/D^{\perp}_d$ could depend on this quantity, and it would if the longitudinal contribution had the standard form $D^{||}_d = v_F^2 \tau/d$. 
However this does not happen for Dirac fermions at neutrality and one instead gets the universal ratio of Eq.~\eqref{eq:53}. 

Let us now quote the final results for the diffusion constant and DC conductivity in physical dimensions $d = 2, 3$. 
We have 
\beq
\label{eq:54}
D_2 = \frac{\gamma^2 \tau}{2 \pi}.
\eeq
According to Eq.~\eqref{eq:53}, one quarter of $D_2$ comes from the band velocity while three quarters from the quantum metric. 
Note that $D_2$ only depends on the Fermi velocity implicitly, through the momentum relaxation time $\tau$, unlike the diffusion constant of a regular metal, which 
is explicitly proportional to $v_F^2$. 

Just as before, we may obtain the DC conductivity from the Einstein relation $\sigma_d = e^2 g_d D_d$. 
The density of states, induced at the Dirac point by disorder, is given by
\beqa
\label{eq:55}
g_d&=&- \frac{1}{\pi} \sum_s \int \frac{d^d k}{(2 \pi)^d} \textrm{Im} G^R_s(\bk, 0) \nonumber \\
&=&\frac{1}{\pi \tau} \int \frac{d^d k}{(2 \pi)^d} \frac{1}{\epsilon_\bk^2 + \frac{1}{4 \tau^2}} = \frac{1}{\pi \gamma^2 \tau}. 
\eeqa
This gives
\beq
\label{eq:56} 
\sigma_2 = \frac{e^2}{\pi \gamma^2 \tau} \frac{\gamma^2 \tau}{2 \pi} = \frac{e^2}{2 \pi^2} = \frac{e^2}{\pi h}, 
\eeq
which is the well-known result for the DC conductivity of Dirac fermions at charge neutrality~\cite{Ludwig94,Ziegler97,Ando98,Shovkovy02,CastroNeto06,Mirlin06}.
The conductivity is much smaller than that of a good 2D metal and is ``universal", i.e. independent of both the Fermi velocity and the strength of disorder. 
As we have shown above, three quarters of this universal conductivity can be related to the quantum geometric tensor, although there is no way to separate
the band velocity from the quantum geometric contributions experimentally, since they only differ by a numerical coefficient. 

For 3D Dirac/Weyl semimetals we obtain 
\beq
\label{eq:57}
D_3 = \frac{\gamma^2}{8 \pi v_F}, 
\eeq
which, according to Eq.~\eqref{eq:53}, comes entirely from the quantum geometric term. 
This gives the following result for the DC conductivity
\beq
\label{eq:58}
\sigma_3 = e^2 g_3 D_3 = e^2 \frac{1}{\pi \gamma^2 \tau} \frac{\gamma^2}{8 \pi v_F} = \frac{e^2}{8 \pi^2 \ell} = \frac{e^2}{4 \pi h \ell}, 
\eeq
which is consistent with previous results~\cite{Fradkin86-1,Fradkin86-2,Altland16}. 
Again, it is interesting that a Dirac fermion, while being in a sense the exact opposite of a flat band (zero effective mass rather than infinite), nevertheless shares the same property that DC transport arises microscopically entirely from quantum geometry rather than band velocity. 
\section{Discussion and conclusions}
\label{sec:3}
In this paper we have demonstrated that the diffusion constant of Dirac fermions in any dimension $d \geq 2$ admits a natural decomposition into band velocity and 
geometric contributions, which is directly related to the well-known separation of a rank two tensor into longitudinal and transverse parts. 
This provides a clearer picture than the one based on direct Kubo formula calculations of the DC conductivity, where these distinct contributions are not as cleanly separated. 
This way of discussing the DC transport problem also reveals a connection to the quantum geometric contributions to superconductivity and magnetism, since the diffusion 
constant may be viewed as the stiffness of the Goldstone modes, associated with spontaneous breaking of the retarded/advanced symmetry. 

We have found that, in the case of 2D Dirac fermions at charge neutrality, exactly one quarter of the diffusion constant, and therefore the DC conductivity, may be related to the standard average of the band velocity squared, while the remaining three quarters are due to quantum geometry. 
Interestingly, in the case of 3D Dirac fermions, the entire result is due to quantum geometry, since the band velocity contribution vanishes exactly (within SCBA and for Gaussian-distributed disorder). 
We find that this exact cancellation of the band velocity contribution is special to 3D, but otherwise accidental, i.e. not related (at least not in any obvious way) to physical properties of 3D Dirac fermions. 

This is interesting, since we typically think of diffusion as an essentially classical process, in which particles with random velocity directions execute a random walk. 
This random walk in the presence of a density gradient is what leads to diffusive equilibration. This is true even in such highly quantum mechanical systems as ordinary metals.
However, this intuition apparently fails in metallic systems without a Fermi surface, in particular Dirac/Weyl semimetals, where diffusion has to do mostly, or even exclusively,
as in the 3D case, with how wavefunctions at different momenta overlap. 
Macroscopically, the diffusive relaxation process is still the same, but the microscopic mechanism is different, which manifests in an entirely different dependence of the diffusion 
constant on various microscopic parameters. 

It would be interesting to extend these considerations to systems without full rotational symmetry, which is only an approximate symmetry of the low-energy Dirac fermion models, considered here. 
Without the full rotational symmetry, the simple transverse-longitudinal tensor decomposition loses its meaning, but it should still be possible to separate band velocity 
from geometric contributions using the fact that the band velocity is always normal to constant energy surfaces in momentum space. The transverse-longitudinal decomposition 
may then be replaced with the decomposition into tangent and normal space components with respect to constant energy surfaces.

\begin{acknowledgments}
Financial support was provided by the Natural Sciences and Engineering Research Council (NSERC) of Canada (computing and manuscript preparation) and the Center for Advancement of Topological Semimetals, an Energy Frontier Research Center funded by the U.S. Department of Energy Office of Science, Office of Basic Energy Sciences, through the Ames Laboratory under contract DE-AC02-07CH11358. 
Research at Perimeter Institute is supported in part by the Government of Canada through the Department of Innovation, Science and Economic Development and by the Province of Ontario through the Ministry of Economic Development, Job Creation and Trade.
\end{acknowledgments}
\bibliography{references}
\end{document}